%Paper: hep-ph/9308355
%From: "Krishna Rajagopal" <krishna@puhep1.Princeton.EDU>
%Date: Sat, 28 Aug 93 13:51:35 EDT

\input phyzzx
\nonstopmode
\twelvepoint
\nopubblock
\overfullrule=0pt
\tolerance=5000
\sequentialequations

\line{\hfill }
\line{\hfill August, 1993}

\titlepage

\title{Emergence of long wavelength pion oscillations following
        a rapid QCD phase transition}

\author{Krishna Rajagopal\foot{Address after Sept. 15:   Department of Physics,
        Harvard University, Cambridge, MA 02138. \hfil\break
        Talk given at the Quark Matter '93 conference in Borl\"ange,
        Sweden, in June, 1993.\hfil\break
        Research supported by a Charlotte Elizabeth Procter
                Fellowship.}}

\vskip .2cm
\centerline{{\it Department of Physics }}
\centerline{{\it Joseph Henry Laboratories }}
\centerline{{\it Princeton University }}
\centerline{{\it Princeton, N.J. 08544 }}

\abstract{
To model the dynamics of the chiral order parameter in a far from
equilibrium QCD phase transition, we consider quenching in the
O(4) linear sigma model.  We summarize arguments and numerical
evidence which show that in the period immediately following
the quench arbitrarily long wavelength modes of the pion field are
amplified.  This results in large regions of coherent pion oscillations,
and could lead to dramatic phenomenological consequences in
ultra-relativistic heavy ion collisions.

My talk was a description of work done with Frank Wilczek [1].
In these proceedings, I sketch our
central results, emphasizing several points that were raised
in discussions at the conference.
The interested reader
should, however, consult Ref. [1] and our earlier
work [2] for a more detailed exposition.
}

\chapter{Misalignment of the chiral condensate}
Among the most interesting speculations regarding ultra-high energy
hadronic or heavy nucleus collisions is the idea that
regions of misaligned vacuum might occur [3].
%If we parametrize the chiral condensate using the usual
%variables of the sigma model,
Misaligned regions are places
where the four-component field $\phi^\alpha
\equiv (\sigma ,\vec \pi )$, that in the
ground state takes the value $(v,0)$ is instead partially
aligned in the $\vec \pi$ directions.  Because of the explicit
chiral symmetry breaking ({\it i.e}
because the pion is not massless), in such a region $\phi$
would oscillate about the $\sigma$ direction.  If
they were produced, misaligned vacuum regions would relax
by coherent pion emission --- they would
produce clusters of pions bunched in
rapidity with highly non-Gaussian charge distributions.
In each such cluster, the ratio
$$
R \equiv {n_{\pi^0} \over n_{\pi^0}+n_{\pi^+}+n_{\pi^-}}
\eqn\ratio
$$
is fixed.  Among
different clusters,
%(in different
%regions in the same collision or in different collisions)
$R$ varies and is distributed according to
$$
{\cal P} (R)= {1\over 2}R^{-1/2}~.
\eqn\probdist
$$
As an example of (1.2), we note that the probability that the
neutral pion fraction $R$ is
less than .01 is .1!  This is a graphic illustration of how different
(1.2) is from what one would expect if individual pions
were emitted with no isospin correlations
many pions.
We have proposed [1] a concrete mechanism by which such
phenomena may arise in heavy ion collisions for which the
plasma is far from thermal equilibrium.

\chapter{Emergence of long wavelength pion oscillations following a quench}

In studying the behaviour of the plasma in the central rapidity region
of a heavy ion collision at RHIC energies or higher,
it seems reasonable to assume that after a time of order 1 fm
a hot plasma is formed in which the chiral order parameter is disordered
and in which the baryon number density is low enough that it can
be neglected.  Our goal is to study the behaviour of
the long wavelength modes of the chiral order parameter as this
plasma loses energy and $\sigma$ develops an expectation value.

In previous work (Ref. [2], references therein, and Wilczek's
talk at this conference) we considered the equilibrium
phase structure of QCD.  We argued that QCD with two massless
quark  flavours probably undergoes a second order transition.
At first sight, this might seem ideal for the development of
large regions of misaligned vacuum, since the long wavelength
critical fluctuations characteristic of a second order transition
are such regions.  Unfortunately, the effect of the light quark
masses spoil this possibility [2].
While in lattice simulations it is in principle possible
to reduce the light quark masses below their  physical values and get
arbitrarily close to the second order critical point,
in heavy ion experiments we must live with a pion which has
a mass comparable to the transition temperature.
Near $T_c$, the correlation length
in the pion channel is shorter than $T_c^{-1}$ [2], and
as a result the misaligned regions
%are
%modest affairs with sizes comparable to the thermal wavelength,
%and
almost certainly do not contain sufficient energy to radiate
large numbers of pions.

Here, we consider an idealization which is in some ways opposite to that
of thermal equilibrium, that is the occurence of a sudden quench
from high to low temperatures, in which the $(\sigma,\vec \pi)$
fields are suddenly removed from contact with a high temperature
heat bath and subsequently evolve mechanically according to
zero temperature equations of motion.
In a real heavy ion collision, the phase transition
proceeds by a process in between an equilibrium phase transition
in which the temperature decreases arbitrarily slowly and a quench in which
thermal fluctuation ceases instanteously.
Our goal is not to quantitatively model a realistic heavy ion collision as
a quench.
Rather, in studying the dynamics of the long wavelength
modes of the chiral order parameter in a quench, our hope is that
the qualitative behaviour in this model is representative of
the physics which occurs in real processes in which the QCD plasma
cools rapidly and is far from thermal equilibrium.

We use the linear sigma model
to describe the low energy interactions of pions:
$$
{\cal L} = \int {\rm d}^4x \left\{ {1\over 2} \partial^\mu \phi^\alpha
\partial_\mu \phi_\alpha ~-~ {\lambda\over 4} ( \phi^\alpha \phi_\alpha
- v^2 )^2 ~+~ H\sigma \right\} ~,
\eqn\lagrangian
$$
where $\lambda$, $v$, and $H\propto m_q$ are to be thought of as
parameters in the low energy effective theory obtained after
integrating out heavy degrees of freedom.  We treat (2.1) as it stands
as a classical field theory, since the phenomenon we are attempting
to describe is basically classical and because as a practical matter it would
be prohibitively difficult to do better.

Our numerical simulations of
quenching in the linear sigma model are described in more detail in [1].
As initial conditions,
we choose $\phi$ and $\dot \phi$ randomly independently on each site
of a cubic lattice.  Therefore, the lattice spacing $a$ represents
the correlation length in the disordered initial state.
In [1] we
made a crude attempt to choose initial distributions
for $\phi$ and $\dot \phi$ appropriate for a quench from an initial
temperature of $T= 1.2 T_c$.  With initial conditions chosen,
we then model the $T=0$ evolution of the system after the quench
by evolving the initial configuration using a standard finite difference,
staggered leapfrog scheme
according to the equations of motion
obtained by varying (2.1).
After each two time steps, we compute the spatial fourier transform
of the configuration and from that obtain the angular averaged
power spectrum.

The central result of our simulations is that the power in the
long wavelength modes of the pion field grows dramatically.
While the initial power spectrum is white and while at late times
the system is approaching a configuration in which the energy
is partitioned equally among modes, at intermediate times of order
several times $m_\pi^{-1}$ the low momentum pion modes are oscillating
with large amplitudes.  When we used gaussian initial distributions
for $\phi$ and $\dot\phi$ with width $v/2$ and $v$ respectively,
the power in modes with $k = 0.2 a^{-1} \simeq 0.3 m_\pi$
is more than 1000 times that in the initial white power spectrum.
For
initial conditions
chosen to model an initial temperature of $1.2 T_c$, the
amplification is less, but is still of order 100.
%In our simulations, after a quench long wavelength
%oscillations of the pion field are amplified.

The amplification of low momentum modes which we observe
in the numerical simulations
can be qualitatively understood by approximating $\phi^\alpha
\phi_\alpha (\vec x,t)$ in the equations of motion by
its spatial average.  After doing the spatial fourier transform,
the equation of motion for the pion field becomes
$$
{d^2 \over dt^2} \vec \pi (\vec k,t) =
- \{ - \lambda v^2 + \lambda \langle \phi^2 \rangle (t)  + k^2 \} \vec \pi
(\vec k,t)
\eqn\eom
$$
where $\langle \phi^2 \rangle (t)$ means simply the spatial average of
$\phi^2$. At late times, $\langle \phi^2 \rangle$
becomes time independent and takes its vacuum value, and
the quantity in brace brackets in (2.2)
becomes simply $m_\pi^2 + k^2$.  Immediately after the quench, however,
$\langle \phi^2 \rangle$ varies with time, and there are periods
when $\langle \phi^2 \rangle < v^2$.  A wave vector $k$ mode of the
pion field is unstable and grows exponentially whenever
$\langle \phi^2 \rangle < v^2 - k^2/\lambda$.   As
$\langle \phi^2 \rangle$ varies, longer wavelength modes are
unstable for more and for longer intervals of time, and, in agreement
with the numerical simulations, are amplified relative to
shorter wavelength modes.

\chapter{Charge separation does not occur}

The striking prediction (1.2) for the probability distribution of
the neutral pion fraction $R$ naturally leads to the question
of whether there are similarly unusual fluctuations in the
electric charge itself, {\it i.e.} in the ratio of $\pi^+$
to $\pi^-$ mesons.
%The charged pions are not related
%to the neutral one by an isospin transformation.
Formulae similar to (1.2) hold for the real fields
$\pi^1 = {1\over \sqrt{2}}(\pi^+
+ \pi^-)$ and $\pi^2 = {1\over i\sqrt{2}}(\pi^+ - \pi^-)$,
but not for $\pi^+$ and $\pi^-$.
While the total electric charge must be conserved, there is no
conservation law prohibiting the separation of charge
into regions of net positive and negative charge.
We must determine whether long wavelength oscillations of
the electric charge density grow. The charge operator
%\begin{equation}
$j_0 = \pi^1 {\partial\over \partial t} \pi^2 -
 \pi^2 {\partial\over \partial t} \pi^1 $
%\end{equation}
measures rotary motion in the $\pi^1 - \pi^2$ plane.  However,
the amplification mechanism which operates following a
quench kicks $\vec \pi$ radially outward, and does
not induce rotary motion.  This heuristic argument is borne
out in the simulations.  Long wavelength oscillations of the
electric charge density are not amplified.

Notice that the question of whether charge separation occurs is
a dynamical one, and has a straightforward
dynamical answer for the quench mechanism of generating regions
of misaligned condensate.  In earlier work,
Kowalski and Taylor [3] imposed isospin symmetry
by hand in order to avoid the possibility of charge separation,
which they consider physically implausible.  The coherent states
we reach are not isospin singlets, and we see no reason
to impose that  condition; nevertheless the intuition of
Kowalski and Taylor is vindicated and there is no charge separation.

\chapter{How large are the regions of coherent pion oscillations?}

%How large are the regions of coherent pion oscillations?
This question is of
crucial phenomenological interest.
In order to be observable,
these regions must evolve into
sufficiently many pions.
We must ask, therefore, what are the longest wavelength modes of the
pion field which get amplified?
In our simulations, the answer is unequivocal --- the wavelength
of modes which are amplified is limited only by the size of
the lattice on which the simulation is run.
Alas, it is much harder to determine what will happen in a real
heavy ion collision.  The one thing that can be said with certainty
is that the effect is {\it not} cut off by the inverse pion
mass.  Modes with $k<m_\pi$ {\it are} amplified.  This is
in marked contrast to the
situation obtained in thermal equilibrium, and is why the phenomenon
%under discussion
will only be detected if the plasma in a heavy ion collision
is far enough from thermal equilibrium that quenching is an
appropriate idealization.  If $m_\pi^{-1}$ is not the
long wavelength cut-off, what is?  The most optimistic
(and perhaps implausible) possibility,
which is suggested by a literal interpretation of
our simulations, is that coherent oscillations of the pion
field in regions as large as the transverse extent of the plasma
are possible.
(The rapid expansion in the longitudinal
direction will damp the growth of modes with $\vec k$ parallel
to the beam
relative to those with $\vec k$ transverse.)
In a real collision, the size of modes which grow could
perhaps be limited by
the time available
before the pions no longer interact and therefore
can no longer be described by oscillations of a classical field,
or perhaps by
the size of regions of
the plasma in which the energy density is reasonably homogeneous.
%%One way of thinking about this \cite{Kajantie}
%%is to realize that
%In our
%simulations, we quenched the entire lattice at the same time.
%In a real collision, the transition process
%will occur later in regions with higher energy density
%than in regions with lower energy density.
%Clearly, work remains to be done.
At this point, the most that can be
said
is that long wavelength pion oscillations
are amplified after a quench, and that their size is not
limited by
any microphysical length like $m_\pi^{-1}$ but is
limited dynamically, perhaps only by the system size.

\chapter{Outlook}

Although we have made many idealizations and approximations,
it seems possible that the essential qualitative feature of the
phenomenon we have elucidated --- long wavelength pion modes
experiencing periods of negative mass$^2$ and consequent
growth following a quench --- could occur in real heavy ion
collisions.  Given the explicit symmetry breaking which
gives mass to the pions, one might have expected the dynamics
following a quench to be featureless.  The mechanism here
discussed provides a robust counterexample.
We have not come close to modelling a real heavy ion collision.
While our treatment can surely be improved, it seems doubtful that
quantitative theoretical predictions for heavy ion collisions
will be possible.  At the end of
the day, the question of whether or not long wavelength pion
oscillations occur will be answered experimentally.
If a heavy ion collision is energetic enough  that there
is a central rapidity region of high energy density and low baryon number,
and if such a region cools rapidly enough that the process
can be modelled as a quench, this will be detected by observing
clusters of pions of similar rapidity in which the neutral pion
fraction $R$ is fixed.  This ratio will be different in different clusters
and will follow a distribution like (1.2).  Were such a signature
to be observed experimentally, it would be clear evidence
for an out of equilibrium transition from a
QCD plasma in which
the chiral order parameter was initially disordered.

\REF\quench{K. Rajagopal and F. Wilczek,
to appear in Nucl. Phys. B, hep-ph/9303281,
Princeton preprint PUPT-1389, IASSNS-HEP-93/16, 1993.}

\REF\eqbm{K. Rajagopal and F. Wilczek, Nucl. Phys. B399 (1993) 395.}

\REF\others{A. Anselm and M. Ryskin, Phys. Letters B226 (1991) 482;
J.-P. Blaizot and A. Krzywicki, Phys. Rev. D46 (1992) 246; J. D. Bjorken,
Int. J. Mod. Phys. A7 (1992) 4189; J. D. Bjorken, Acta Phys. Pol. B23 (1992)
561; K. L. Kowalski and C. C. Taylor, preprint hepph/9211282, 1992;
J. D. Bjorken, K. L. Kowalski, and C. C. Taylor, SLAC preprint
SLAC-PUB-6109, 1993.}

\refout

\end